\documentclass[aps,amsmath,amssymb,showpacs,prb,floatfix,twocolumn]{revtex4}
\usepackage{amsmath}

\usepackage{graphicx}

\usepackage{epsfig}

\newcommand \bea {\begin{eqnarray} }

\newcommand \eea {\end{eqnarray}}

\newcommand{\ccmo}{Ca$_3$Co$_{2-x}$Mn$_x$O$_6$ }

\newcommand{\ccco}{Ca$_3$Co$_2$O$_6$ }

\newcommand{\uudd}{$\uparrow \uparrow \downarrow \downarrow$}
\newcommand{\uuud}{$\uparrow \uparrow \uparrow \downarrow$}
\newcommand{\uuuu}{$\uparrow \uparrow \uparrow \uparrow $}

\newlength{\figwidth}
\newlength{\shift}
\newcommand{\fg}[3]
{
\begin{figure}[h]
\vspace*{-0cm}
\[
\includegraphics[width=\figwidth]{#1}
\]
\vskip -0.2cm
\caption{\label{#2}
\small#3
}
\end{figure}}
\begin{document}

\title{Analysis of magnetization and a spin state crossover in the multiferroic \ccmo
}

\author{R. Flint}
\email{flint@physics.rutgers.edu}
\author{H.-T. Yi}
\author{P. Chandra}
\author{S.-W. Cheong}
\author{V. Kiryukhin}

\affiliation{
Department of Physics and Astronomy and Center for Emergent Materials,
Rutgers University, Piscataway, NJ 08855, U.S.A. 
}

\pacs{77.80.-e,75.10.Pq,75.80.+q,71.70.-d,75.30.Wx}

\begin{abstract}

\ccmo ($x \sim 0.96$) is a multiferroic
with spin-chains of alternating Co$^{2+}$ and Mn$^{4+}$ ions.
The spin state of Co$^{2+}$ remains unresolved,
due to a discrepancy between high temperature  
X-ray absorption ($S=\frac{3}{2}$) and low temperature neutron ($S=\frac{1}{2}$) measurements.  
Using a combination of magnetic modeling and crystal-field analysis, we show 
that the existing low temperature data cannot be reconciled within
a high spin scenario by invoking spin-orbit or Jahn-Teller distortions.
To unify the experimental results, we propose
a spin-state crossover with specific experimental predictions.
\end{abstract}

\date{\today}

\maketitle

Multiferroics where the spin-lattice coupling arises
from symmetric superexchange
offer great promise for large magnetoelectric effects\cite{Mostovoy07}.
\ccmo($x\sim0.96$) (CCMO)
is one such material\cite{Zubkov01,Rayaprol03,Mostovoy07,Choi08,Jo09,Wu09,Zhang09,Yao09,Kiryukin09,Lancaster09},
consisting of anisotropic spin chains where the \uudd \: 
ordering of alternating Co$^{2+}$ and 
Mn$^{4+}$ spins breaks inversion symmetry.  Understanding the magnetic properties is key to understanding the multiferroicity;
however, the spin state of the Co$^{2+}$ ions remains unresolved.  Here we address this problem directly.
High temperature (T) susceptibility ($\chi$) \cite{Zubkov01} and X-ray absorption ($T \sim 300 K$)
spectroscopy (XAS) measurements \cite{Wu09}, along with first principles calculations,\cite{Wu09,Zhang09}
indicate it is
in a high spin (HS) ($S = \frac{3}{2}$) state
with a large orbital moment ($1.7 \mu_B$,)
whereas low T neutron studies \cite{Choi08} are 
consistent with $S = \frac{1}{2}$.
In this paper, we demonstrate that the measured high-field 
magnetization\cite{Jo09} ($m(H)$) confirms
the small moment at low T, and new $\chi$
measurements indicate that the moment anisotropy
increases with decreasing temperature.
One might think that the Co ion remains in the HS state,
but that its effective spin is reduced at low T.
Here we use a combination of magnetic modeling and crystal-field (CF) analysis
to show that no perturbation of the $S=3/2$ states can account for {\sl both} the observed moment
and the large anisotropy at low T, thereby ruling out this simple picture.
In order to resolve the experimental situation, we propose a spin-state crossover for the Co$^{2+}$ ion.  Low T XAS 
measurements would test our proposal directly, and we also predict a number of other experimental 
consequences.

CCMO consists of c-axis chains of Co$^{2+}$ and 
Mn$^{4+}$ ions, arranged in a triangular lattice\cite{Choi08}.  
These atoms reside in alternating octahedral and trigonal prismatic 
oxygen cages; Mn preferentially
sits on the octahedral sites.   
Magnetic \uudd \: order develops along the chains at $T^* = 16K$.
Exchange striction causes alternating bonds to constrict and lengthen.  As the two ions have different charges, a macroscopic c-axis 
polarization results\cite{Choi08}.  
Comparison of the $\chi$ 
parallel and perpendicular 
to the chains indicates that these spins 
have an Ising anisotropy that increases with decreasing T(see Fig. 1).

\fg{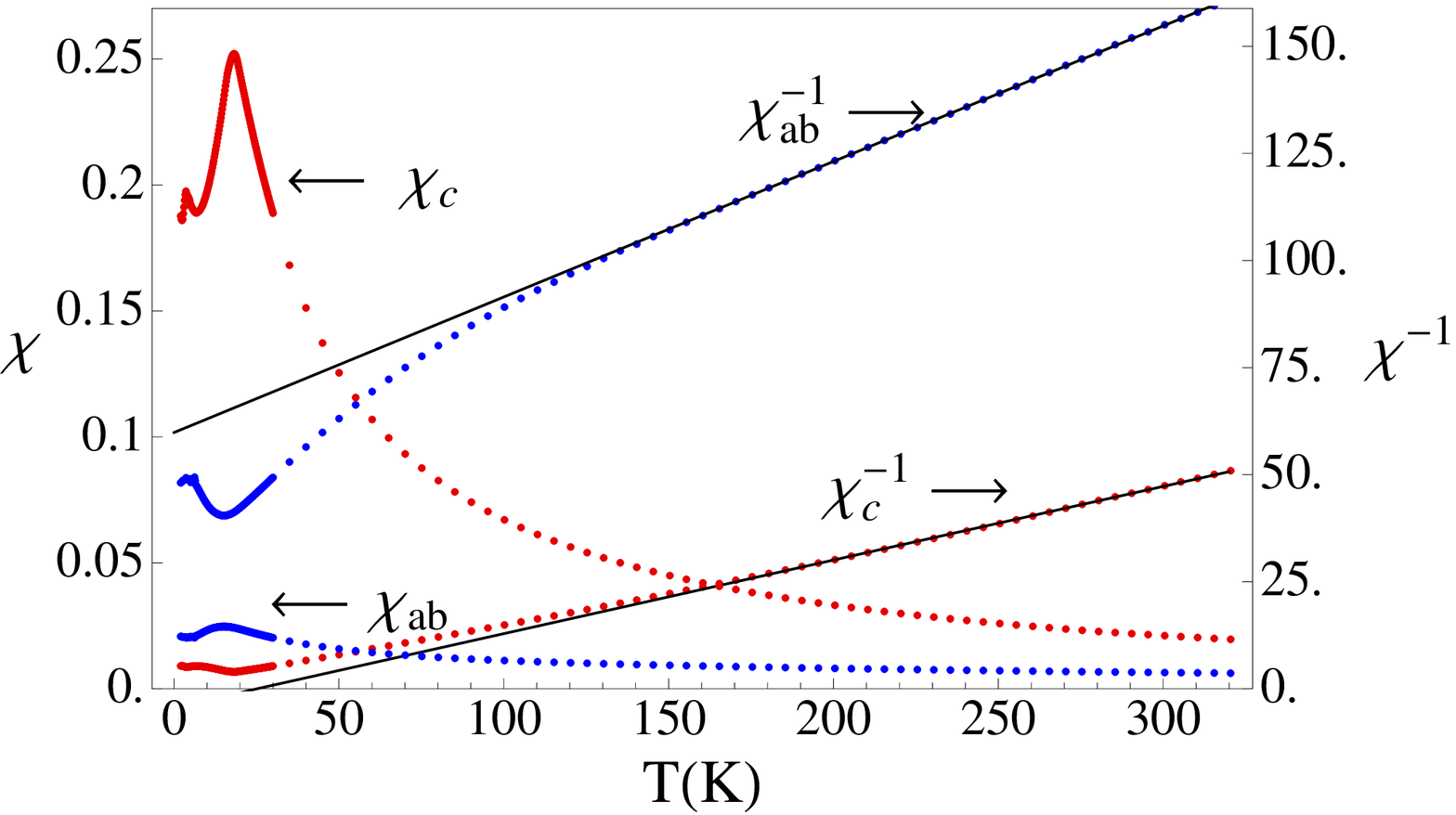}{susceptibility}{Susceptibility along the a- and c-axes, indicating increasing Ising anisotropy with decreasing temperature.  Sample preparation described in Choi et al\cite{Choi08}.}

We now turn to the spin states of the magnetic ions.  
Mn$^{4+}$ is a $3d^3$ ion in an octahedral
environment.  The $t_{2g}$ orbital is 
half-filled, quenching the angular momentum; thus
Mn$^{4+}$ acts as a $S=\frac{3}{2}$ Heisenberg spin, consistent
with both low-T neutrons\cite{Choi08} and high-T 
XAS\cite{Wu09}.  
This spin state is stable against both spin-orbit (SO) 
coupling and structural distortions  up to the CF splitting $\sim 1$ eV.

Thus both the anisotropy and the small moment of CCMO at low T
must be due to the Co. Co$^{2+}$ is a $3d^7$ ion with three holes in 
a trigonal prismatic environment.  The CF levels, shown in Fig. \ref{ss1}A, 
are labeled by their $L_z$ quantum numbers \cite{dai}. 
While Hund's rules and the CFs are compatible 
in Mn$^{4+}$, here they compete.
Hund's rules align the holes to yield $S = \frac{3}{2}$, 
while the CF energy is reduced by placing all three holes in the lowest level to get $S = \frac{1}{2}$.  Both scenarios have 
partially filled levels 
resulting in orbital moments and anisotropy.
We must thus rely on experiment to discern the spin state  
of Co$^{2+}$. Curie-Weiss fits to $\chi$, for $T \gg T^*$ yield an effective moment of $\mu_{eff} \sim 6.0 \mu_B$ 
indicating that both Mn$^{4+}$ and Co$^{2+}$ are in the high-spin (HS)
$S=\frac{3}{2}$ states\cite{Zubkov01,Rayaprol03}, consistent with   
room T XAS studies\cite{Wu09}.  
However the moment measured by neutron scattering for 
$T < T^*$,
$0.66 \mu_B$/Co, is significantly less than that expected for the HS scenario, suggesting
a LS state\cite{Choi08}. 
There is a clear discrepancy between these results, and 
we turn to another low T  measurement, $m(H)$, to resolve this 
issue.

\fg{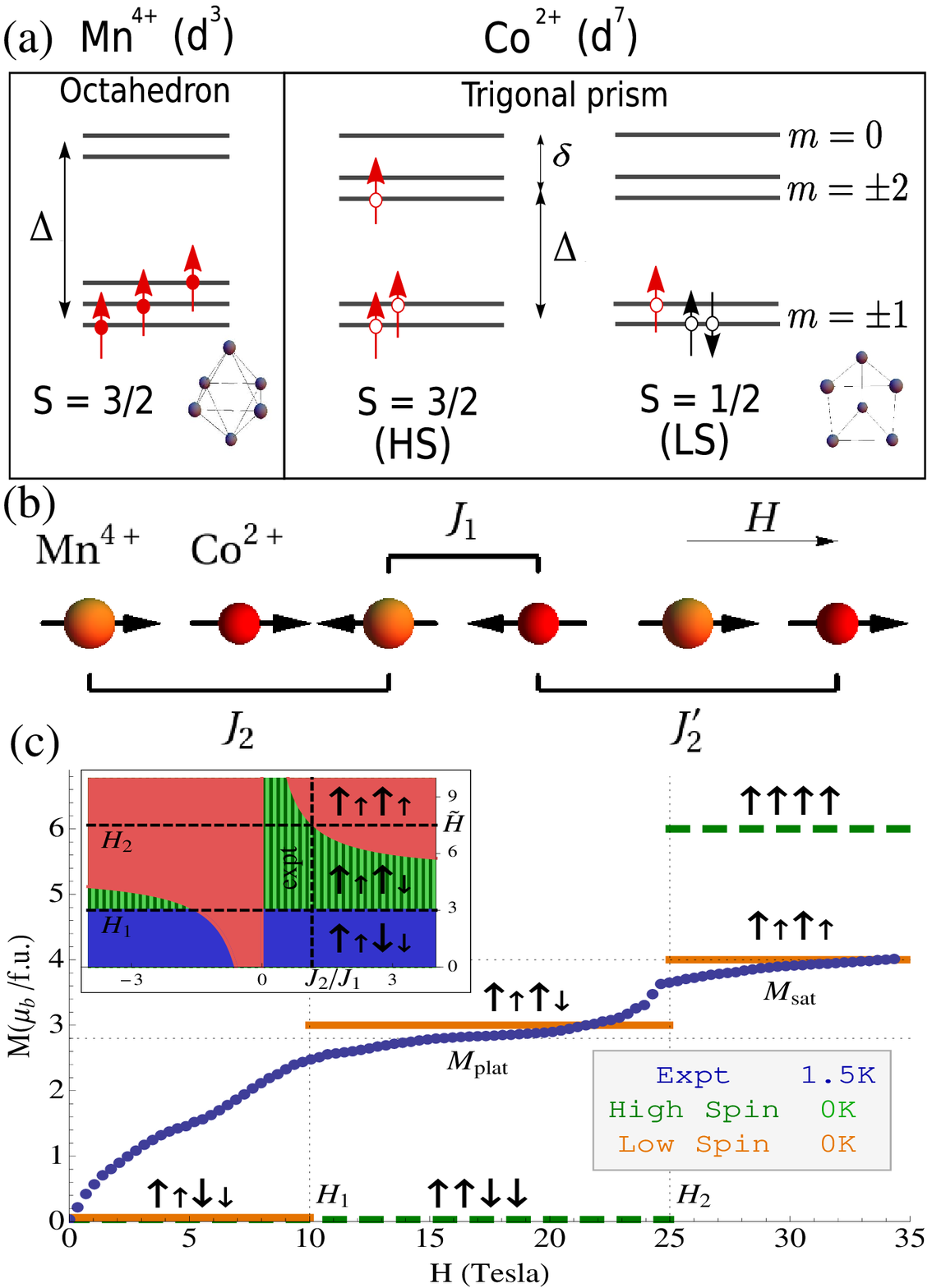}{ss1}{(a) Crystal field levels and occupations for (Left) Mn$^{4+}$ in an octahedral environment 
where the $t_{2g}$ level is half-filled; the orbital angular momentum is 
thus quenched and Mn$^{4+}$ has a $S=\frac{3}{2}$ Heisenberg spin.  
(Right) Co$^{2+}$ is
simply described by three holes, with the crystal fields inverted from 
the electron picture.  In the trigonal environment, there is a competition between the crystal 
field splitting and the Hund's rule coupling. Both high- and low-spin 
ground states have unquenched angular momentum($\langle L_z \rangle$ = 2 and 1 respectively) and 
are possible. (b) Magnetic structure of the Co$^{2+}$-Mn$^{4+}$ chains, showing the in-chain couplings. (c) 
High field magnetization data (at 1.5K) reproduced from \cite{Jo09}, 
overlaid with the LS and HS Co cases at $T=0$. (Inset)Example $T=0$ phase diagram of the alternating spin ANNNI model($s = \frac{1}{2}, S=\frac{3}{2}$, $J_2'/J_2 = 5$), 
where $\tilde{H} = \frac{H}{g\mu_B J_2}$ and $J_2/J_1$ are varied.
For $3 < J_2'/J_2 < 6$, the experimental ratio $H_2/H_1 \sim 2.5$ requires $J_1 > 0$, 
while for $J_2'/J_2 > 6$, $J_1 < 0$.  Any first principles calculations must satisfy these relations.}

In Fig. \ref{ss1}(b), we show the 
spin structure of the chains, and, in Fig. \ref{ss1}(c) 
reproduce the $m(H)$ data
published elsewhere\cite{Jo09}.  $m(H)$ has 
two steps, the first with a height of $2.8 \mu_B/f.u.$ and 
the second, $4 \mu_B/f.u.$.  
As T decreases, the spins align in the 
\uudd\: ordering\cite{Choi08}, and
an anisotropic next-nearest neighbor Ising (ANNNI) model\cite{Bak82} 
with alternating spins\cite{pini,Kiryukin09,Kim96} has successfully described
many properties of CCMO\cite{Yao09}; here we apply it to $m(H)$ to resolve the spin-states.  The Hamiltonian is
\begin{equation}
\label{ccmoH}
\mathcal{H} = \sum_{i}J_1 s_i S_i + J_2 S_i S_{i+1} + J_2' s_{i} s_{i+1} - 
\mu_B H_z(g_{s} s_i+g_{S}S_i ),
\end{equation}
where 
$s_i$ and $S_i$ refer to Co and Mn spins in cell $i$ respectively. 
The exchange couplings are given in Fig. \ref{ss1}(b), where 
$J_2$ and $J_2'$ are not identical; 
the sign of $J_1$ cannot be reliably determined from high T $\chi$\cite{Haverkort08},
and we find that both signs can reproduce $m(H)$(Fig. \ref{ss1}(c) inset).

When analyzing $m(H)$ for $H || c$, it is  
useful to discuss moments, $\mu_z(Co,Mn)$, not spins and g-factors,
where $\mu_z(Mn) = g_S S = 3$, and 
$\mu_z(Co) = g_s s$ is unknown.  
For $\mu_z(Co) < \mu_z(Mn)$, the ground states of (\ref{ccmoH}) 
go from \uudd \: to \uuud \: to \uuuu \: as $H_z$ increases, 
yielding the $m(H)$ steps shown in Fig. \ref{ss1}(c); 
the ratio of plateau heights is
\begin{equation}
\frac{M_{plat}}{M_{sat}} = \frac{\mu_z(Mn)}{\mu_z(Mn)+\mu_z(Co)} = 
\frac{1}{1+\frac{\mu_z(Co)}{\mu_z(Mn)}} = \frac{3}{4},
\end{equation}
where $3/4$ is the experimental value from Fig \ref{ss1}c.
The linear behavior and hysteresis\cite{Jo09} below the first step are due to polarization domain walls, which are free spins\cite{domain_note}.

The presence of two steps in $m(H)$ demands two different moments.   
These plateaus can be easily explained within the LS scenario; here, as $H$ 
increases it first flips the 
large  (Mn) spin (at $H_1$) and later (at $H_2> H_1$$)$  the small (Co) 
one.  The first step then  
has a value of $\mu_z(Mn) = 3 \mu_B$; 
this is consistent
with  the measured value of $2.8 \mu_B$; and
neutron scattering has confirmed that the first plateau is the expected 
\uuud\: state\cite{Jo09}. 
The height of the second step gives 
$\mu_z(Co) = g_{Co} S_{Co} = 1.2 \mu_B$, 
where $g_{Co} \geq 2$ due to the nonzero orbital moment;
this is larger than that observed in neutron 
scattering\cite{Choi08} but three times smaller than that 
associated with $S=\frac{3}{2}$.  
The HS scenario would instead give a second plateau 
height between $5.8-7.8\mu_B$, depending on the orbital moment,
and the simple HS picture is inconsistent with $m(H)$.  
For completeness, we note that the final step can be explained with a HS Co
by a tripling of the magnetic unit cell at high fields; this unlikely scenario requires a failure of the ANNNI model,
but could be checked by neutron diffraction. 
Dimensional fluctuations have also been suggested to reduce the HS Co moment to its observed value\cite{Wu09}, but the equally large Mn moment is only
suppressed by $5\%$.
The combination of the neutron scattering and $m(H)$ data strongly implies a small Co moment.

We examine the theoretical situation of the Co ion more carefully;
by using general considerations
we can rule out the HS scenario and show that the LS scenario can account for both the small moment and large anisotropy.
We treat only the lowest CF levels, $|L_z = \pm m,S_z\rangle$, assuming that these are well separated, and introduce
both SO and Jahn-Teller(JT) as perturbations,
\begin{equation}
\mathcal{H} = -|\lambda|\vec{L}\cdot\vec{S} - \mu_B(\vec{L}+2\vec{S}) \cdot \vec{H} - \delta JT (E_a - E_b),
\end{equation} 
where the negative SO term comes from the alignment of spin and orbital moments in hole ions.  The JT term,
$\delta JT (E_a - E_b)$ splits $|L_z = \pm m\rangle$ into $|a,b\rangle=\frac{1}{\sqrt{2}}\left( |m\rangle \pm |-m\rangle\right)$,
quenching the angular momentum.

First we try to obtain a small effective moment at low T within the HS scenario; here
the Hund's energy, $J_H$ is larger than the CF splitting, 
$\Delta$ (see Fig. \ref{ss1}), and the ``true'' spin is $\frac{3}{2}$.  
The lowest ($L_z = \pm 1$) level is half-filled; however the 
$L_z = \pm 2$ level is partially occupied, leading to an orbital 
moment of $2 \mu_B$.  The ground-state is eightfold 
degenerate, and is split into four Kramers doublets by either SO
or JT interactions.
With only SO splitting,
$J_z = L_z + S_z$ remains a good quantum number, and the 
larger $J_z$ levels lie lowest(Fig. \ref{ss2}A).  The ground state 
doublet has $J_z = \pm \frac{7}{2}$ with a moment $\mu_z = L_z + 2 S_z = 5 \mu_B$, 
while $\mu_x = 0$.  Thermal mixing lowers the anisotropy, 
causing $\chi_c/\chi_a$ to increase with decreasing T; however,
$\mu_z$ decreases only slightly with T and is 
four times larger than the observed moment at 
$T \sim  |\lambda| \approx \lambda_0/2S = 230K$ in HS Co\cite{fazekas}.

Clearly SO splitting cannot explain the small moment, but what about JT distortions, which
will quench the orbital contribution?
Though JT is a small effect, if present, in CCMO\cite{Zhang09},
we consider the extreme case, $\delta JT \gg |\lambda|$ as an example.  
These four-fold degenerate $S=\frac{3}{2}$ levels will be further split by second-order effects(see Fig. \ref{ss2}B).  
The general form of this Hamiltonian is constrained by the 
trigonal ($S_z \rightarrow S_z, S_{\pm} \rightarrow \mathrm{e}^{\pm i 2\pi/3}S_{\pm}$) and time-reversal symmetries 
($\vec S \rightarrow -\vec S$) 
that only allow operators of the form $S_z^2$ or $S_+ S_-$.  If we consider the virtual fluctuations into the other JT quartet, 
the energy shift will be $\Delta E = - |\langle a|H_{SO}|b\rangle|^2/\delta JT = -4\lambda^2 S_z^2/\delta JT$, 
lowering the $S_z = \pm \frac{3}{2}$ level to restore the anisotropy while preserving the spin-only moment, $\mu_z = 3\mu_B$ and $\mu_x = 0$.  
Still, this moment is more than two times too large.  Moreover, this is a ``best-case moment''; both thermal mixing and
stronger SO coupling will only increase it, and these should be quite important.   
Even if we could change the sign of $\Delta E$ to lower the $S_z = \pm \frac{1}{2}$ doublet, 
perhaps by fluctuations into nearby CF levels, its $S = \frac{3}{2}$ origins mean that, while the c-axis moment is small, $\mu_z = 1\mu_B$, the basal 
plane moment, $\mu_x = 2 \mu_B$ is large, the inverse of the observed anisotropy. 

\fg{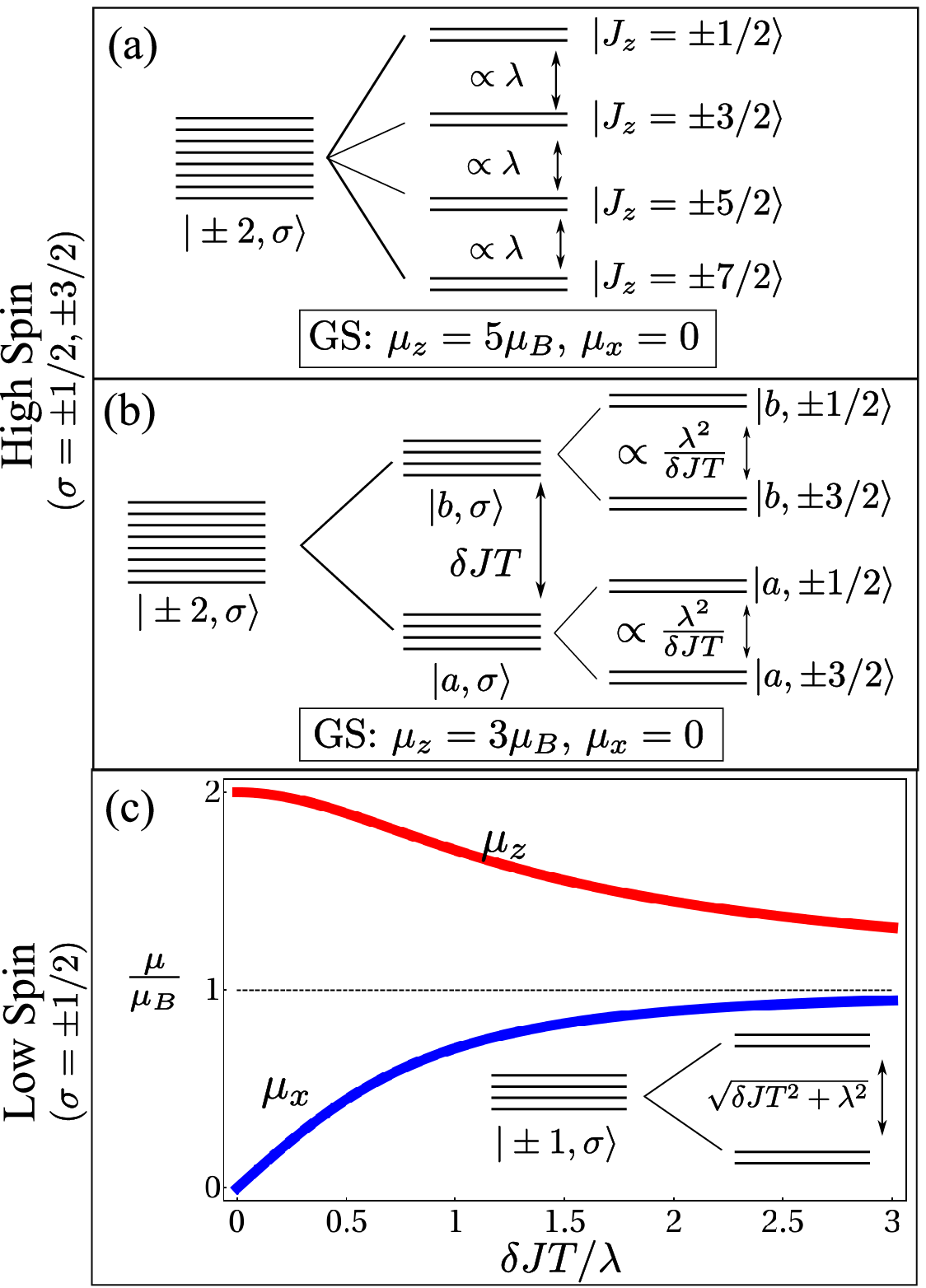}{ss2}{
Spin orbit(SO) and Jahn Teller(JT) distortions reduce both the $S = \frac{3}{2}, Lz = \pm 2$(HS) octet and $S = \frac{1}{2}, Lz = \pm 1$(LS) quartet to Kramer's doublets, but with a range of moments and anisotropies.
(A.) HS SO split states: as Co$^{2+}$ is more than half filled, SO aligns the spin and orbital moments.
(B.) HS JT split states: JT splits $|\pm m\rangle$ into  
$|a,b\rangle \equiv \frac{1}{\sqrt{2}}\left( |m\rangle \pm |-m\rangle\right)$, quenching $\langle \vec{L} \rangle$; but
the anisotropy is restored by second-order SO effects, which favor the $S_z = \pm \frac{3}{2}$ level for small $\delta JT$. 
(C.) For the LS case, we allow $\delta JT/\lambda$ to vary, splitting the quartet into doublets, with the ground state ranging from perfectly anisotropic,
$\mu_z = 2 \mu_B, \mu_x = 0$ for pure SO to an isotropic $S=\frac{1}{2}$ for pure JT.}

None of the possible HS states can simultaneously explain the low moment and Ising anisotropy at low Ts; 
however they describe the 
high T data well.  Therefore we turn to the LS scenario at low T, and propose
a spin-state crossover between 16K and 300K.  To obtain such a crossover, the
difference in energy scales $J_H - \Delta$ must be small and negative; the LS 
ground state is energetically selected, but the
entropy, $S \sim \log(2S+1)$ favors the HS state at higher T.   While SO splits both the HS and LS states into doublets, there are more HS states, closer together(as $\lambda = \lambda_0/2S$).
We note that such spin-state crossovers have been observed before; the best-known example is LaCoO$_3$\cite{Koehler57},
where the $3d^6$ Co$^{3+}$ transitions from a $S=0$ ground state to $S=2$ at 500K\cite{Koehler57}; 
this crossover has been characterized by many probes\cite{naiman65,Haverkort06,Kozlenko07} and serves as a benchmark in our present discussion. 
Co$^{2+}$ spin state transitions have been confirmed in several organic complexes\cite{coII} 
at lower Ts, but never in a crystal.
In these materials the Co ions are in octahedral symmetries, but 
no spin-state transitions have yet been observed in trigonal coordination, where the CF splitting, $\Delta$ 
is generally smaller(see Fig. \ref{ss1});
however density functional theory on the related \ccco indicates that there $\Delta$ is nearly as large as in 
octahedral geometry\cite{Wu05}.  Although first principles calculations on CCMO have found HS Co\cite{Wu09,Zhang09}, 
we believe their $J_H$ may be too large.
Having eliminated the HS scenario, we next pursue the
exotic LS scenario.

The LS state has spin $\frac{1}{2}$, but there is also an unquenched angular momentum, $L_z = \pm 1$ from the
partially occupied level.  
Both SO and JT split this quartet into two Kramers doublets, and we have treated these simultaneously, 
plotting the ground state $\mu_z$ and $\mu_x$ for a range of $\delta JT/\lambda$(Fig. \ref{ss2} C).  
The pure SO case leads to a $J_z = \pm \frac{3}{2}$ doublet, with maximum anisotropy at $T=0$, while
the pure JT case leads to
an isotropic spin $\frac{1}{2}$ at zero T, with increasing anisotropy with $T$.
It is impossible to get both a low moment and a 
large anisotropy from CFs alone in a hole ion; while SO gives rise to the anisotropy, it
also always aligns the orbital and spin moments.

In spin chains, the magnetic fluctuations extend over a large temperature range and cannot be neglected.  As 
magnetism develops along the chains, the spins align and the anisotropy increases; the \uudd\: order also leads to 
exchange striction, causing
a nonzero polarization. It likely also distorts the trigonal environment of the Co, thus acting as a small JT distortion 
to reduce the moment. The concurrent JT reduction of the anisotropy will be small compared to the positive
magnetic contribution, and thus this combination of low-dimensional magnetic fluctuations and exchange striction
will lead to a smaller moment while maintaining the large spin anisotropy.

We therefore propose the following scenario: at the lowest temperatures, Co is in a  spin $\frac{1}{2}$ state, 
where the orbital moment is partially
quenched by exchange striction associated with the developing magnetism.  Entropy favors the spin $\frac{3}{2}$ state 
with a large orbital moment, leading to a spin-state crossover between $T^* = 16K$ and room T.  
This scenario provides a coherent explanation of \emph{all} the experimental evidence, at both low and high T.

The spin state transitions in LaCoO$_3$ are clearly visible as a decrease in the effective moment\cite{naiman65}, while
CCMO features only an weak upturn in 
$\chi_c^{-1}$, seemingly indicating an increase of the moment, and a 
downturn in $\chi_{ab}^{-1}$.  
However, in LaCoO$_3$, the magnetism comes solely from the HS states, while there are several
competing effects in CCMO. The development of magnetism along the chains
increases the effective Co and Mn c-axis moments, causing an upturn in $\chi_c^{-1}$, while the increasing anisotropy of the Mn spin
reduces its ab-plane moment, explaining the downturn in $\chi_{ab}^{-1}$.  
Thermal population of the SO split levels(see Fig. \ref{ss2} A,C) also increases $\mu_z(Co)$ as the temperature decreases, 
with effects down to one third of the SO coupling, near the observed upturn at 150K\cite{Haverkort08}.
Either of these could mask a decreasing moment; moreover, if the Weiss temperature does not reduce with the moment, 
a spin state crossover will manifest as an upturn in $\chi_c^{-1}$
above the crossover temperature followed by a downturn.  For these reasons, the upturn in $\chi_c^{-1}$ does not rule out 
a LS-HS crossover, but it does suggest that the crossover, if it occurs, is at or below 100K.  

The most direct experimental test of the spin-state crossover would be to probe the LS ground state with XAS at low T; 
this method was used to confirm the spin-state crossover in LaCoO$_3$\cite{Haverkort06}.
There are also several other consequences of the spin-state transition.  The excitation between the LS and HS states 
is observable in inelastic neutron scattering. Another magnetization plateau at $5.8-7.8 \mu_B/f.u.$  
will occur at higher fields which favor the HS state.  Finally, as the Co ion volume increases with increasing 
degeneracy\cite{bari72}, a pressure-dependent spin state crossover should occur in CCMO at fixed T.

In conclusion, we have analyzed the field-dependent magnetization and the magnetic anisotropy 
in CCMO to prove that at low temperatures
the Co$^{2+}$ ions are unambiguously in a spin $\frac{1}{2}$ state with a moment consistent with 
earlier neutron studies\cite{Choi08}.
Similarly, high temperature studies\cite{Wu09} indicate that this same ion is unambiguously spin $\frac{3}{2}$.  
We have shown that
this small moment, in combination with the observed anisotropy, cannot be reconciled with a high spin 
scenario by invoking spin-orbit coupling or Jahn-Teller distortions.
We thus propose a low spin-high spin crossover as a function of temperature and suggest a number of 
experimental probes to test this idea.  

We acknowledge helpful discussions with P. Coleman, M. Croft, M. Dzero, M. Haverkort and A. Nevidomskyy.
The susceptibility was measured by Y.J.Choi.
This work was supported by DOE DE-FG02-99ER45790(RF),DOE DE-FG02-07ER46832(HY,SC,VK) and NSF-NIRT-ECS-0608842(PC).


\begin{thebibliography}{99}



\bibitem{Mostovoy07} S-W. Cheong and M. Mostovoy, Nat. Mater. {\bf 6}, 13(2007).



\bibitem{Zubkov01} V.G. Zubkov et al, J. Solid State Chem. {\bf 160}, 293(2001).



\bibitem{Rayaprol03} S. Rayaprol et al., Sol. State Comm. {\bf 128}, 79 (2003).



\bibitem{Choi08} Y.J. Choi et al, Phys. Rev. Lett. {\bf 100}, 047601(2008).



\bibitem{Jo09}Y.J. Jo et al., Phys. Rev. B {\bf 79}, 012407(2009);



\bibitem{Wu09}H. Wu et al, Phys. Rev. Lett. {\bf 102}, 026404 (2009).


\bibitem{Zhang09} Y. Zhang et al.Phys. Rev. B {\bf 79}, 054432(2009).

\bibitem{Yao09} X.Yao and V.C. Lo, J. Appl. Phys. {\bf 106}, 013903(2009).


\bibitem{Kiryukin09}V. Kiryukhin et al., Phys. Rev. Lett. {\bf 102}, 187202(2009).

\bibitem{Lancaster09} T. Lancaster et al., Phys. Rev. B {\bf 80} 020409 (1990).

\bibitem{pini} M.G. Pini and A. Rettori, Phys. Rev. B {\bf 48}, 3240(1993).


\bibitem{Bak82} P. Bak, Rep. Prog. Phys. {\bf 45}, 587(1982).

\bibitem{Kim96} J. Kim et al, J. Phys. Soc. Jap. {\bf 65}, 2624(1996).

\bibitem{dai} D. Dai and M.-H. Whangbo, Inorg. Chem. {\bf 44},12(2005).


\bibitem{Haverkort08} M. W. Haverkort et al,arXiv.org:0806.3736(2008).

\bibitem{domain_note} A polarization domain wall: $\uparrow \uparrow \!\!\underline{\downarrow}\!\! \downarrow\! | \underline{\downarrow}\!\! \uparrow \uparrow \downarrow$ contains two free(E=0) spins(underlined).



\bibitem{fazekas} P. Fazekas, \emph{Lecture Notes on Electron Correlation and Magnetism}(World Scientific,Singapore, 1999)


\bibitem{Koehler57} W.C. Koehler and E.O. Wollan, J. Phys. Chem. Sol. {\bf 2}, 100(1957);

J.B. Goodenough et al., J. Phys. Chem. Sol. 5, 17 (1958).


\bibitem{naiman65} C.S. Naiman et al., J. Appl. Phys. {\bf 36} 1044(1965).


\bibitem{Haverkort06} M. W. Haverkort et al, Phys. Rev. Lett. {\bf 87}, 176405(2006).


\bibitem{Kozlenko07}D.P. Kozlenko et al., Phys. Rev. B 75, 064422 (2007).



\bibitem{coII} H.A. Goodwin, \emph{Topics in Current Chemistry} {\bf 234}, 23(2004).


\bibitem{Wu05}H. Wu et al., Phys. Rev. Lett. {\bf 95}, 186401(2005).

\bibitem{bari72} R. Bari and J. Sivardiere, Phys. Rev. B {\bf 5},4466(1972).





\end{thebibliography}
\end{document}